\documentclass[11pt]{article} 
\usepackage{mystyle-new}
\usepackage{epsfig,amsmath} 
\usepackage{graphics}
\usepackage{hyperref}
\usepackage{cite,./mcite}
\usepackage{listings}
\usepackage{color}
\usepackage{booktabs}
\usepackage{tabularx}
\usepackage{tikz}
\usepackage{longtable} 
  \newcommand{\arrowcom}[1]{\textcolor{red}{\textbf{$\Longrightarrow$ #1}} \\}

\def\??#1{\mbox{}\\\arrowcom{#1}}
 
\pdfoutput=1

  {\begin{list}{}{\settowidth{\labelwidth}{#1}%
  \setlength{\leftmargin}{\labelwidth}%
  \addtolength{\leftmargin}{\labelsep}%
  \setlength{\itemsep}{0pt plus 1pt}
  \setlength{\parsep}{0pt plus 1pt}
  \setlength{\topsep}{0pt plus 1pt}
  \setlength{\partopsep}{0pt plus 1pt}
  \setlength{\parskip}{2mm plus 1mm minus 1mm}
  }}%
  {\end{list}}

\definecolor{listinggray}{gray}{0.9}
\definecolor{lbcolor}{rgb}{0.9,0.9,0.9}
\def\lsim{\mathrel{\rlap{\lower4pt\hbox{\hskip1pt$\sim$}}
    \raise1pt\hbox{$<$}}}                
\def\gsim{\mathrel{\rlap{\lower4pt\hbox{\hskip1pt$\sim$}}
    \raise1pt\hbox{$>$}}}                

\newcommand{\as}{\alpha_\mathrm{s}}

\def\kt{\ensuremath{k_t}}

\newcommand{\CCFM}{CCFMa,CCFMb,Catani:1989sg,CCFMd}

\def\tmdlib{{TMDlib}}  
\def\tmdplotter{{TMDplotter}}

\begin{document}

\begin{flushright}
DESY 14-059 \\
NIKHEF 2014-024 \\
RAL-P-2014-009 \\
YITP-SB-14-24 \\
 Dec 2014
\end{flushright}

\begin{center} {\sffamily\Large\bfseries \tmdlib\  and \tmdplotter : \\
library and plotting tools for \\
transverse-momentum-dependent parton distributions  }
\\ \vspace{0.5cm}

{ \Large
F.~Hautmann$^{1,2}$, 
H.~Jung$^{3,4}$,
M.~Kr\"amer$^{3}$, \\
P. J.~Mulders$^{5,6}$,
E. R.~Nocera$^{7}$,
T. C.~Rogers$^{8,9}$,
A.~Signori$^{5,6}$
} \\
\vspace*{0.15cm}
\end{center}
%
{\large $^1$ Rutherford Appleton Laboratory, UK} \\ 
{\large $^2$ Dept.\  of  Theoretical Physics, University of Oxford,   UK}  \\
{\large $^3$ DESY, Hamburg, FRG}\\ 
{\large $^4$ University of Antwerp,  Belgium}\\ 
{\large $^5$ Department of Physics and Astronomy, VU University Amsterdam, the Netherlands}\\
{\large$^6$ Nikhef, the Netherlands}\\
{\large$^7$ Universit\`a degli Studi di Genova and INFN Genova, Italy} \\
{\large$^8$ C.N. Yang Institute for Theoretical Physics, Stony Brook University, USA} \\
{\large$^9$ Department of Physics, Southern Methodist University, Dallas, Texas 75275, USA}


\begin{abstract}

Transverse-momentum-dependent distributions (TMDs)  
are extensions of collinear parton distributions and are 
important  in high-energy physics 
from both theoretical and phenomenological points of view.
In this manual we introduce the library \tmdlib , 
a tool to collect
transverse-momentum-dependent parton distribution functions (TMD PDFs) 
and fragmentation functions (TMD FFs) together with an online plotting tool, 
\tmdplotter. 
We provide a description of the program components and of the different 
physical frameworks 
the user can access via the available parameterisations.
\end{abstract} 

\newpage

\section*{PROGRAM SUMMARY}
\label{sec:summary}
{\em Computer for which the program is designed and others on which it is operable:}   any with standard C++, tested on Linux and Mac OSX systems \\ \\
{\em Programming Language used:}  C++ \\ \\
{\em High-speed storage required:}  No \\ \\
{\em Separate documentation available: } No \\ \\
{\em Keywords: } QCD, TMD factorisation, high-energy factorisation, TMD PDFs, TMD FFs, unintegrated PDFs, small-$x$ physics.\\ \\
{\em Other programs used:}  LHAPDF (version 6) for access to collinear parton distributions, {\sc Boost} (required by LHAPDF version 6), {\sc Root} (version higher than 5.30) for plotting the results \\ \\
{\em Download of the program:} \verb+http://tmdlib.hepforge.org+ \\ \\
{\em Unusual features of the program:}   None \\ \\
{\em Contacts:}   H. Jung (hannes.jung@desy.de), E. Nocera (emanuele.nocera@edu.unige.it), A. Signori (asignori@nikhef.nl)  \\ \\
{\em Citation policy:} please cite the current version of the manual and the paper(s) related to the parameterisation(s). \\
\newpage

\section{Introduction}
\label{sec:Introduction}

The Quantum Chromodynamics (QCD) interpretation of high-energy particle reactions 
requires a simultaneous treatment of processes at different energy scales. 
Factorisation theorems provide the mathematical framework to properly separate 
the physical regimes. 
For instance, when two protons collide in a Drell-Yan (DY) event
the high-energy partonic cross section is described
with a  perturbative QCD expansion and the soft physics underlying the structure of the hadrons is treated
with parton distribution functions (PDFs), supplemented by QCD evolution. 
``Evolution", in this context, refers to the scale dependence of parton distributions (and similar 
non-perturbative objects) that arises in a detailed treatment of factorisation in QCD perturbation theory.
A classic example of a consequence of QCD evolution is the violation of Bjorken-scaling in inclusive deep-inelastic lepton-hadron scattering (DIS), predicted by the 
Dokshitzer-Gribov-Lipatov-Altarelli-Parisi (DGLAP) evolution equations~\cite{Gribov:1972ri,Altarelli:1977zs,Dokshitzer:1977sg}.

The same basic picture applies to other (semi-)inclusive processes,
like semi-inclusive DIS (SIDIS), and
$e^+e^-$ annihilation into hadrons.
A PDF describes the likelihood for finding a parton of a particular momentum inside an incoming hadron.
In processes with observed hadrons in the final state, fragmentation functions (FFs)
 enter to describe the transition from a partonic state to an observed final-state hadron.

For sufficiently inclusive processes, only the component of parton momentum collinear 
to the momentum of its parent hadron is relevant at leading power ({\it leading twist}) in the hard scale.
Factorisation theorems for such processes are traditionally called {\it collinear} factorisation theorems. 
In less inclusive processes, however, sensitivity to the  partonic motion transverse to the direction 
of the parent hadron can become important. 
In such cases, the PDFs and FFs must carry information about 
transverse parton momentum in addition to the collinear momentum. One must introduce 
transverse-momentum-dependent (TMD) PDFs and FFs and use them in the context of new factorisation theorems, called 
TMD factorisation theorems. 
TMD factorisation has been formulated 
for a number of semi-inclusive processes including SIDIS, DY and $e^+e^-$ 
annihilation~\cite{Collins:1981uk,Collins:1981uw,Collins:1982wa,Collins:1981tt,Collins:1984kg,Collins:2011zzd,Meng:1995yn,Nadolsky:1999kb,Nadolsky:2000ky,Ji:2004wu,Ji:2004xq,GarciaEchevarria:2011rb,Chiu:2011qc}.
For particular processes in hadronic collisions, like heavy flavour or heavy boson (including Higgs) production, 
TMD factorisation has also been formulated in the high-energy (small-$x$) 
limit~\cite{Catani:1990xk,Levin:1991ry,Collins:1991ty,
Hautmann:2002tu}. 
In this context, the functions encoding the hadronic structure are more often 
referred to as 
{\em unintegrated} parton distribution functions (uPDFs), 
see {\it e.g.} Refs.~\cite{Avsar:2012hj,Avsar:2011tz,Jadach:2009gm,Dominguez:2011saa,Dominguez:2011br,Dominguez:2011gc,Hautmann:2009zzb,Hautmann:2012pf,Hautmann:2007gw}.

The presence of a large variety of TMD factorisation and evolution frameworks 
complicates efforts to compare different TMD PDFs/FFs and uPDFs parameterisations. 
In some cases, the differences arise because different formalisms employ similar 
TMD concepts, but are tailored to specific physical applications. An example is the difference 
between the Collins-Soper-Sterman (CSS) style of TMD factorisation discussed 
in Sec.~\ref{TMDfactorisation} compared with the high-energy TMD factorisation style discussed 
in Sec.~\ref{HEfactorisation}.  The former 
is designed for semi-inclusive processes differential 
in a particular physical transverse momentum and with a finite and non-zero ratio between the hard 
scale and the overall energy. The latter (high-energy TMD factorisation) is designed for the limit of 
a fixed hard scale and very high energies. Moreover, within each category there are also 
competing subcategories of approaches. For instances, the detailed phenomenological methods 
that employ a CSS-style of approach in 
Refs.~\cite{Landry:2002ix,Konychev:2005iy,Bozzi:2005wk,Echevarria:2012pw,Guzzi:2013aja,Anselmino:2013lza,Aidala:2014hva,Su:2014wpa} 
are rather different.

In this paper, we describe a new tool for collecting different  
fits and parameterisations 
into a single library, \tmdlib, and the online plotter tool, \tmdplotter. 
Provided that the user takes into account all the possible differences
between formalisms, collecting parameterisations 
for both the objects in \tmdlib\ and \tmdplotter\ will also make phenomenological comparisons easier.  

The paper is organised as follows: In Sec.~\ref{sec:theory}, we briefly 
introduce  the theoretical framework for 
both TMD and high-energy factorisation and evolution. In Sec.~\ref{sec:TMDlibdocumentation},
we present a concise documentation of the {\tmdlib} library and {\tmdplotter} tool, 
discussing the basic procedure to readily use them.

\section{Theoretical framework}
\label{sec:theory}

In this section, we briefly 
describe two different commonly-used frameworks for factorisation and
evolution of parton distributions. 
Specifically, we discuss TMD and high-energy 
factorisation theorems and evolution equations.

\subsection{TMD factorisation and evolution}
\label{TMDfactorisation}

When one hard scale enters a high-energy process (like the invariant mass of the exchanged virtual photon in DIS) 
and the relevant transverse momenta are integrated over, 
one applies {\em collinear} factorisation 
to separate the hard partonic physics from the soft hadronic physics.
When sensitivity to intrinsic transverse momentum is important, 
one must go beyond the collinear framework 
to factorise perturbative and non-perturbative dynamics.
For example, this is the case 
in processes with observed transverse momenta in the final states, 
like SIDIS and DY lepton pair production at low transverse momentum. 
In these cases the low transverse momentum 
provides greater access to novel QCD dynamics 
as compared to the collinear case.
If the observable transverse momenta are much larger than $\Lambda_{\rm QCD}$, then 
often the cross section may be expressed entirely in collinear factorisation, though supplemented by transverse 
momentum resummation. 

Feynman rules allow for a decomposition of the cross section 
into a contraction of hadronic and leptonic tensors. 
Where applicable, factorisation theorems separate  
non-perturbative and hard contributions within the hadronic tensor. 
In the TMD case, distribution and fragmentation functions are introduced, whose 
properties depend on the polarisations of the target and/or produced hadrons, 
the partonic polarisations, and the twist order.
For example, in fully unpolarised SIDIS at leading twist the hadronic tensor is factorised 
into a convolution of one 
unpolarised TMD PDF (for the incoming target hadron) and one unpolarised TMD FF (for the final state hadron):
\begin{multline}
\label{e:had_tens}
W^{\mu \nu} \sim {\cal H}^{\mu \nu}(Q;\mu) \sum_a \int d^2 {\bf b}_{\perp} e^{-i {\bf q}_{\perp} 
\cdot {\bf b}_{\perp}} f^{a,T} (x,{\bf b}_{\perp};\zeta_f,\mu)\ D^{a \to h} (z,{\bf b}_{\perp};\zeta_D,\mu)\ + \\ + Y_{\rm SIDIS}({\bf q}_\perp,Q) + {\cal O}((\Lambda_{\rm QCD}/Q)^{p})\ ,
\end{multline}
where ${\cal H}$ is the hard part, $a$ is the flavour of the struck parton, $T$ is the target hadron, 
$h$ is the detected hadron, $x$ and $z$ are the light-cone momentum fractions, 
and ${\bf b}_{\perp}$ is the
Fourier conjugate of the transverse momentum ${\bf q}_{\perp}$. 
The function $f^{a,T} (x,{\bf b}_{\perp};\zeta_f,\mu)$ is a TMD PDF while $D^{a \to h} (z,{\bf b}_{\perp};\zeta_D,\mu)$ is a TMD FF. 
The scale $\mu$ is a renormalization group scale, $\zeta_{f,D}$ are rapidity evolution scales. 
$Q$ is the hard scale that enters into the hard vertex. 
In SIDIS $Q = \sqrt{-q^2}$, where $q$ is the four-momentum of the exchanged virtual photon.

The term $Y_{\rm SIDIS}({\bf q}_\perp,Q)$ is a correction for the region of $q_\perp \sim Q$ where 
a separation into TMDs is not valid, and all transverse momentum is generated inside the hard scattering.
This so-called $Y$-{\it term} is calculable in collinear factorisation. With it included, the corrections are
suppressed by powers of $\Lambda_{\rm QCD}/Q$, point-by-point in ${\bf q}_\perp$, as indicated by the last term, where $p > 0$.
Taking into account all the possible combinations of polarisation (parton, target and detected hadron), there are eight TMD PDFs and eight TMD FFs at leading-twist, although the number of operator combinations could be larger~\cite{Buffing:2012sz,Buffing:2013eka}. The expression 
of the hadronic tensor is modified accordingly~\cite{Mulders:1995dh,Boer:1997nt,Bacchetta:2006tn}.

TMD parton distributions or fragmentation functions depend on two types of
auxiliary scales, $\zeta_{f,D}$ and $\mu$, and they satisfy 
evolution equations with respect to both of them. 
The evolution with respect to $\zeta_{f}$ and 
$\zeta_D$ corresponds to Collins-Soper (CS) evolution and is determined by a process-independent 
soft factor~\cite{Collins:1999dz,Collins:2004nx,Konychev:2005iy,Hautmann:2007uw,Collins:2011zzd,GarciaEchevarria:2011rb,Cherednikov:2008ua,Cherednikov:2009wk,Chiu:2012ir,Echevarria:2014rua}. 
The scales $\zeta_f,\ \zeta_D$ must satisfy the constraint $\zeta_f \zeta_D = Q^4$.
The evolution in $\mu$, instead, is determined by standard renormalisation group methods.

When the energy range covered by the experimental data is not large (see, e.g., Ref.~\cite{Airapetian:2012ki,Adolph:2013stb}) fits of TMD PDFs and FFs can be performed without taking into account effects induced by evolution.
These fits rely essentialy on a simple parton model approach and are oriented towards investigations of hadron structure at a relatively low-energy scale. Recent examples are Refs.~\cite{Signori:2013mda,Anselmino:2013lza}.
In order to explore the evolution of hadron structure with the energy scale, these fixed scale fits can be incorporated into a Collins-Soper-Sterman (CSS) style of factorisation theorem like Eq.~\eqref{e:had_tens}, as described 
in Refs.~\cite{Aybat:2011zv,Aybat:2011ge}. There, fixed scale fits 
from~\cite{Schweitzer:2010tt,Anselmino:2005an,Anselmino:2008sga,Collins:2005ie,Collins:2005rq} 
are combined with traditional CSS style fits from Refs.~\cite{Landry:2002ix,Konychev:2005iy}.

\subsection{High-energy factorisation and evolution}
\label{HEfactorisation}

A form of TMD factorisation holds at high 
energy~\cite{Catani:1990xk,Catani:1990eg,Catani:1993ww}  and has been 
applied to several processes  
in photon-hadron, lepton-hadron and hadron-hadron 
collisions.  For instance, the 
high-energy factorisation   
expresses  
the heavy-quark  leptoproduction cross section in terms of the TMD gluon 
density via well-prescribed, calculable perturbative coefficients~\cite{Catani:1990eg}. 
This framework is extended to deep-inelastic  structure functions  in Refs.~\cite{Catani:1994sq,Catani:1993rn}. 
Perturbative applications of the method include the resummation 
of small-$x$ logarithmic corrections to DIS to  all orders in $\alpha_s$  
at leading and next-to-leading $\ln x$ 
level~\cite{Catani:1994sq,Catani:1993rn,Ciafaloni:1998gs,Fadin:1998py}.
In hadron-hadron scattering, 
high-energy factorisation has been applied to processes such as 
heavy flavour  and Higgs  boson  production~\cite{Catani:1990eg,Hautmann:2002tu}. 

In the framework of high-energy factorisation~\cite{Catani:1990xk,Catani:1990eg,Catani:1993ww} 
the DIS cross section can be written as a convolution in both longitudinal and transverse momenta  
of the unintegrated parton density function 
${\cal  A}\left(x,\kt,\mu\right)$    
with off-shell partonic matrix elements
\begin{equation}
 \sigma_j  ( x , Q^2 )  = \int_x^1  
d z  \int d^2k_t \ \hat{\sigma}_j( x   ,  Q^2 ,  {    z}   ,  k_t ) \ {\cal  A}\left( { z} ,\kt,  \mu \right)  , 
\label{kt-factorisation}
\end{equation}
where  the DIS cross sections 
$\sigma_j$, ($j= 2 , L$) 
are related to the  structure functions $F_2$ and $F_L$ 
 by 
$\sigma_j = 4 \pi^2 F_j / Q^2$, and   
the hard-scattering kernels ${\hat \sigma}_j$ of Eq.~(\ref{kt-factorisation}) 
are $k_t$-dependent.  

The factorisation formula, Eq.~(\ref{kt-factorisation}), allows for resummation of logarithmically 
enhanced $x\to 0 $ contributions to all orders in perturbation theory, both in the hard-scattering coefficients 
and in the parton evolution, taking into account the full dependence on the factorisation scale $\mu$ and 
on the factorisation scheme~\cite{Catani:1994sq,Catani:1993rn}.  

Realistic applications of this approach at collider energies require 
matching of $x \to 0$ contributions with  finite-$x$ contributions. 
To this end, the evolution of the gluon uPDF ${\cal  A} $ is obtained by combining the resummation of  
small-$x$ logarithmic contributions~\cite{Lipatov:1996ts,Fadin:1975cb,Balitsky:1978ic}   
with medium- and large-$x$ contributions to parton splitting~\cite{Gribov:1972ri,Altarelli:1977zs,Dokshitzer:1977sg}, 
according to the CCFM evolution equations~\cite{\CCFM}. 

The cross section $\sigma_j$ ($j= 2 , L$) is usually computed in a Fixed Flavour Number (FFN) scheme, 
where  the photon-gluon fusion process ($\gamma^* g^* \to q \bar{q}$) is included. 
The masses of the  quarks are explicitly included with the light and heavy quark masses being free parameters.
In addition to $\gamma^* g^* \to q\bar{q}$, the contribution from valence quarks is included via 
$\gamma^* q \to q$ by using CCFM evolution of valence 
quarks~\cite{Deak:2010gk,Deak:2011ga,Hautmann:2013tba}. 
A fit of CCFM uPDFs to the combined DIS 
precision data~\cite{Aaron:2009aa,Abramowicz:1900rp}    
has been recently presented in Ref.~\cite{Hautmann:2013tba} using the evolution   given in Ref.~\cite{Hautmann:2014uua}.
Earlier CCFM fits to DIS were  presented in Ref.~\cite{jung-dis04}. 
In Ref.~\cite{wuesthoff_golec-biernat} the unintegrated gluon distribution has been obtained by means of a saturation ansatz.

\newpage

\section{{\tmdlib} documentation}
\label{sec:TMDlibdocumentation}

{\tmdlib} is a {\tt C++} library which provides a framework and an  interface 
to a collection of different uPDF/TMD parameterisations.  The parameterisations of TMDs in \tmdlib\ are explicitly authorised for distribution in \tmdlib\ by the authors.
No explicit QCD evolution code is included: the parameterizations are as given in the corresponding references. 
In the present version of \tmdlib\ no attempt is made to unify grid files and the interpolation procedure, both are those provided by the authors. 

The source code of {\tmdlib} is available from 
{\url{http://tmdlib.hepforge.org/}~} 
and can be installed using the \textit{standard} {\tt autotools} sequence 
configure, make, make install, with options to specify 
the installation path and the location of the
{\tt LHAPDF} PDF library~\cite{LHAPDF6,web:LHAPDF} 
and the {\tt ROOT} data analysis framework library~\cite{Brun:1997pa,web:ROOT} (which is used optionally for plotting).
If {\tt ROOT} is not found via {\tt root-config}, the plotting option is disabled.
After installation, {\tt TMDlib-config} gives access to necessary environment variables.

The up-to-date list of all the available functions can be found at \url{http://tmdlib.hepforge.org/namespaceTMDlib.html}, and is also summarized in Tabs.~\ref{tab:TMDinit}-\ref{tab:TMDpdf}-\ref{tab:tmdlibmethods}.
The {\tmdlib} calling sequence is:  {\bf Initialisation} (selecting the desired uPDFs/TMDs), see Tab.~\ref{tab:TMDinit};
{\bf Call} (producing the uPDF/TMD for partons at $x$, $\mu$ and $k_\perp$), see Tab.~\ref{tab:TMDpdf};
{\bf Information} (displaying details about the initialised uPDFs/TMDs), see Tab.~\ref{tab:tmdlibmethods}. 
Note that function overloading is used to create different methods for the functions devoted to uPDF/TMD initialisation ({\tt TMDinit}) and  call ({\tt TMDpdf}).

\begin{itemize}
\item {\bf INITIALISATION. } 
The first step consists in initialising the desired uPDF/TMD set.
Initialisation assigns the chosen uPDF/TMD set, specified by its name, 
an identifying number proper to that set.\footnote{Note that only 
one set of uPDF/TMD at a time can be called so far.} 
This number is stored into memory
and called each time the identification of the uPDF/TMD set is needed by any 
{\tmdlib} internal function. The complete list of uPDF/TMD sets available in 
is given in Tab.~\ref{tab:TMD/uPDF_sets} with the corresponding name, 
identifier, kinematic coverage, and reference. 
This list will be constantly updated at
\url{http://tmdlib.hepforge.org/pdfsets.html}
as soon as new uPDF/TMD sets will become available.

The TMD fit of Ref.~\cite{Signori:2013mda} is provided as a Monte Carlo
ensemble of $N_{\mathrm{rep}}=200$ equally probable replicas, as both a 
grid with polynomial interpolation and the analytic form with the {\it 
best-fit} parameters for each replica. The user should specify the replica to be
initialised and whether he would like to use the grid or the parameterisation
via the input variables {\tt irep} and {\tt imode} respectively.
Through {\tt imode} it is also possible to select the Fourier transform of the TMD PDF, namely the distribution in transverse coordinate space ($b_T$-distribution).
For other uPDF/TMD sets, these options are not available and, if specified, 
they will be ignored.

\item {\bf CALL TO THE DISTRIBUTION. }
The second step consists in calling the desired function.
Specifically, 
the light-cone momentum fractions $x^+$ and $x^-$ (often set $x^-=0$) carried by the parton, 
the parton transverse momentum $k_t$ (in \textrm{GeV}), 
the energy scale $\mu$ (in \textrm{GeV}) and the flavour code identifying 
the target\footnote{uPDF/TMD parameterisations have been determined 
for proton or antiproton only so far.} are the input variables. Returned is the momentum weighted parton distribution.
\end{itemize}

\begin{table}[!h]
\footnotesize
\centering
\begin{tabular}{clllllcc}
\toprule
Parton    & uPDF/TMD set      & identifier   & $\Lambda^{(4)}_{qcd}$ & $\kt^{cut}$ [GeV]& $Q_0$ [GeV]& Ref. \\ 
\midrule
Gluon     & ccfm-JS-2001      & 101000 & 0.25 & 0.25 & 1.4  & \cite{jung-dis04} \\
          & ccfm-setA0        & 101010 & 0.25 & 1.3  & 1.3  & \cite{jung-dis04} \\
          & ccfm-setA0+       & 101011 & 0.25 & 1.3  & 1.3  & \cite{jung-dis04} \\
          & ccfm-setA0-       & 101012 & 0.25 & 1.3  & 1.3  & \cite{jung-dis04}\\ 
          & ccfm-setA1        & 101013 & 0.25 & 1.3  & 1.3  & \cite{jung-dis04}\\ 
          & ccfm-setB0        & 101020 & 0.25 & 0.25 & 1.3  & \cite{jung-dis04}\\
          & ccfm-setB0+       & 101021 & 0.25 & 0.25 & 1.3  & \cite{jung-dis04}\\
          & ccfm-setB0-       & 101022 & 0.25 & 0.25 & 1.3  & \cite{jung-dis04}\\ 
          & ccfm-setB1        & 101023 & 0.25 & 0.25 & 1.3  & \cite{jung-dis04}\\ 
          & ccfm-JH-set 1     & 101001 & 0.25 & 1.33 & 1.33 & \cite{jung-dis03}\\  
          & ccfm-JH-set 2     & 101002 & 0.25 & 1.18 & 1.18 & \cite{jung-dis03}\\  
          & ccfm-JH-set 3     & 101003 & 0.25 & 1.35 & 1.35 & \cite{jung-dis03}\\  
          & ccfm-JH-2013-set1 & 101201 & 0.2  & 2.2  &  2.2 & \cite{Hautmann:2013tba} \\
          & ccfm-JH-2013-set2 & 101301 & 0.2  & 2.2  &  2.2 & \cite{Hautmann:2013tba} \\
          & GBWlight          & 200001 & --   & --   & --   & \cite{wuesthoff_golec-biernat} \\
          & GBWcharm          & 200002 & --   & --   & --   & \cite{wuesthoff_golec-biernat} \\
          &  KS-2013-linear    & 400001 &  0.3    &  --    &   --    & \cite{Kutak:2012rf} \\
          &  KS-2013-non-linear    & 400002 & 0.35 &  --    &  --    & \cite{Kutak:2012rf} \\
\midrule
Quark     & ccfm-setA0        & --     & 0.25 & 1.3  & 1.3  &  \\
          & ccfm-JH-2013-set1 & --     & 0.2  & 2.2  & 2.2  & \cite{Hautmann:2013tba} \\
          & ccfm-JH-2013-set2 & --     & 0.2  & 2.2  & 2.2  & \cite{Hautmann:2013tba} \\
          & SBRS-2013-TMDPDFs & 300001 & --   & --   & 1.55 & \cite{Signori:2013mda} \\
\bottomrule
\end{tabular}
\caption{Available uPDF/TMD parton sets in \tmdlib .}
\label{tab:TMD/uPDF_sets}
\label{updfs}
\end{table}

Additional methods, utility routines and examples available in {\tmdlib} are:
\begin{itemize}
 \item {\tt TMDutils}: collection of methods used in {\tmdlib},
 including functions to get details about the initialised uPDF/TMD set 
 (like $\alpha_s$, $\Lambda_{\mathrm{QCD}}$, number of flavours),
 see Tab.~\ref{tab:tmdlibmethods};
 \item {\tt TMD$\_$test}: example program to handle uPDF/TMD distributions;
 \item {\tt TMDplotter}: {\tt ROOT}-based script to plot uPDF/TMD 
 distributions as obtained from {\tmdlib}.
\end{itemize}

\begin{figure}[h]
\includegraphics[scale=0.75,width=7.8cm,height=10.0cm]{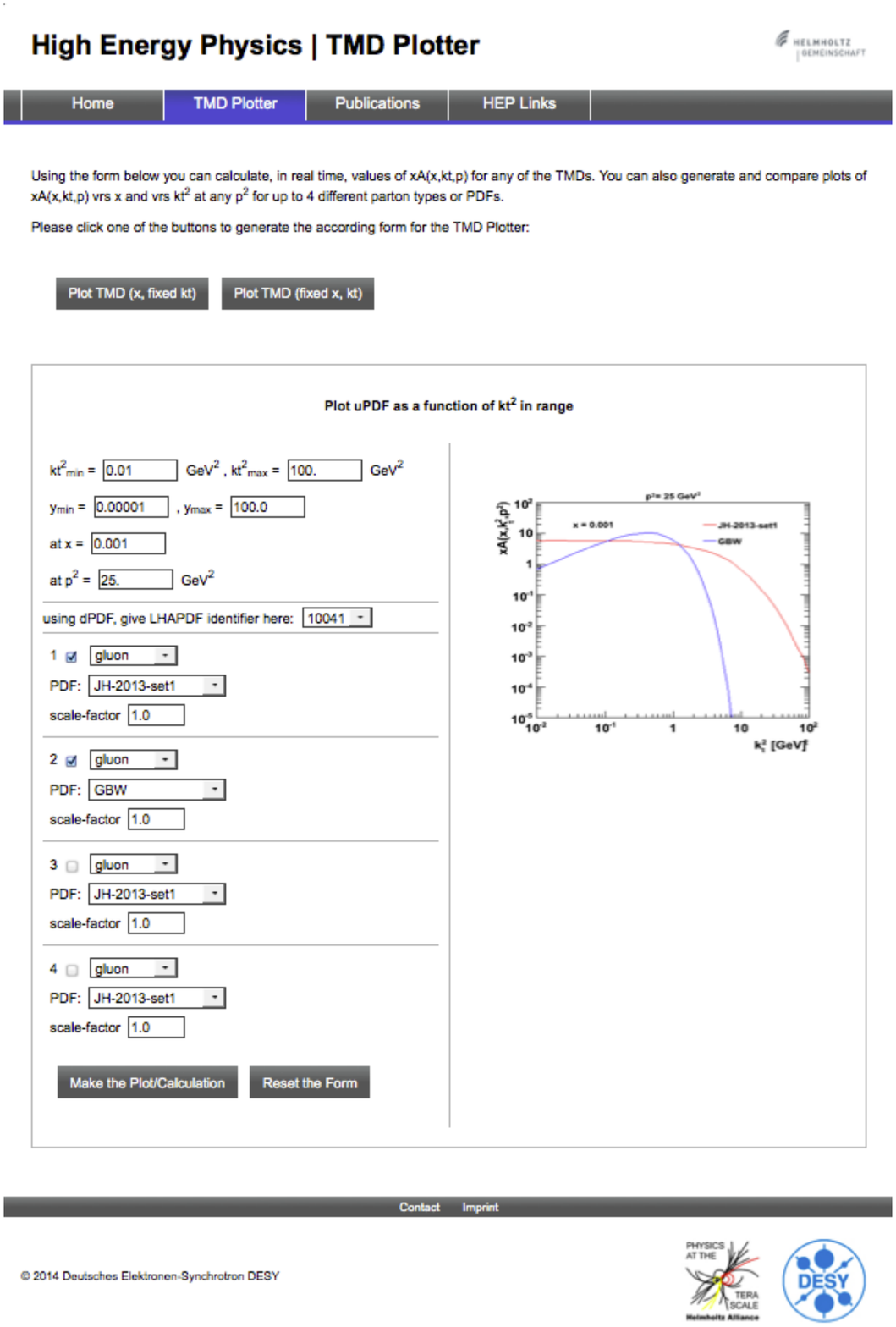}
\vskip -10.6cm \hskip 8.5cm
\includegraphics[scale=0.75,width=7.3cm,height=11.2cm]{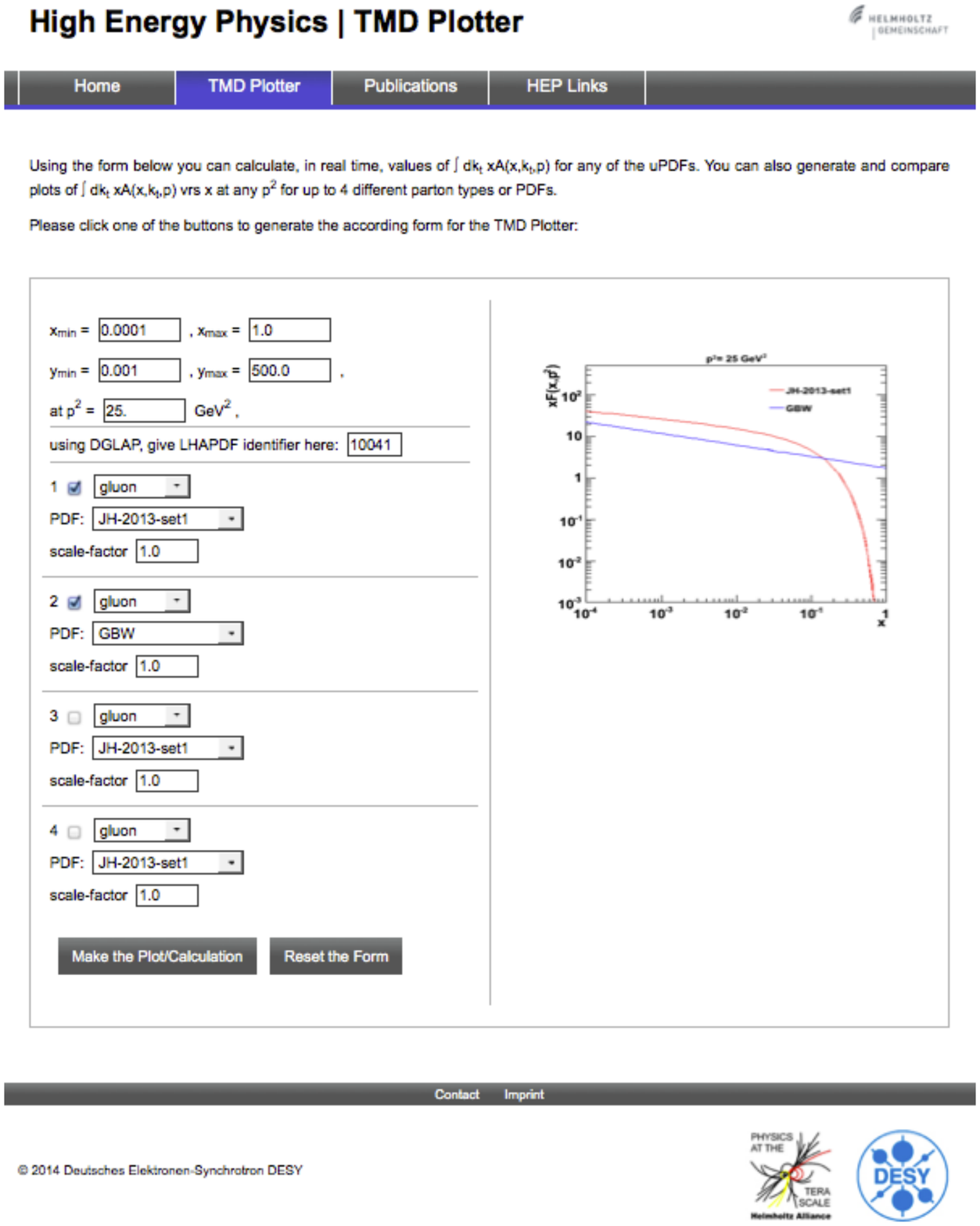} \\
\begin{center}
\caption{Two snapshots from the online portal \tmdplotter\ for plotting uPDF/TMD distributions:
the gluon from the {\tt ccfm-JH-2013-set1} set compared to the 
{\tt GBW} as a function of $k_t$ (left) and $x$ (right).}
\label{fig:snapshot} 
\end{center}
\end{figure}

\begin{table}[!h]
 \footnotesize
 \centering
 \begin{tabularx}{\textwidth}{p{0.3\textwidth}X}
  \toprule
  Method & Usage\\
  \midrule
  {\tt TMDinit(name)} & To initialise the uPDF/TMD set specified by its name {\tt name}.
                      A complete list of uPDF/TMD sets available in the current version of
                      {\tmdlib} with the corresponding name is provided in 
                      Tab.~\ref{tab:TMD/uPDF_sets}.\\
  \midrule
  {\tt TMDinit(name,irep)}
                      & To initialise a given {\tt irep} replica in a Monte Carlo uPDF/TMD 
                      set specified by its name {\tt name}.\\
  \midrule
  {\tt TMDinit(name,irep,imode)}
                      & To initialise the uncertainty sets with {\tt irep} or
                      to initialise a given {\tt irep} replica in a Monte Carlo uPDF/TMD 
                      set specified by its name {\tt name} and {\tt imode}:
                      \begin{itemize}
                        \item {\tt imode}=0: the value obtained from the analytic form of the distribution is returned   
                        \item {\tt imode}=1: the value obtained as a polynomial interpolation on a numerical grid is returned
                        \item {\tt imode}=2: the value obtained from the analytic form of the Fourier transform (b-space distribution)                                          
                      \end{itemize}\\
  \bottomrule
 \end{tabularx}
 \caption{The function overload for {\tt TMDinit} used to initialise uPDF/TMD parton sets.}
 \label{tab:TMDinit}
\end{table} 
\begin{table}[!h]
 \footnotesize
 \centering
 \begin{tabularx}{\textwidth}{p{0.3\textwidth}X}
  \toprule
  Method & Usage\\
  \midrule
  {\tt TMDpdf(x,xbar,kt,mu, uval,dval,s,c,b,glu)}
                       & Void-type function filling the variables {\tt uval}, {\tt dval},
                       {\tt s}, {\tt c}, {\tt b}, {\tt glu} with the values of 
                       $xF(x,\bar{x},k_t,\mu)$ ($F$ is the initialised uPDF/TMD)
                       for valence u-quarks {\tt uval}, valence d-quarks {\tt dval}, light sea-quarks {\tt s}, 
                       charm-quarks {\tt c}, bottom-quarks {\tt b}, and gluons {\tt glu} respectively
                       for a proton target. 
                       The input variables {\tt x} and {\tt xbar} are 
                       the light-come momentum fractions $x^+$ and $x^-$
                       (in some parameterisations the latter is set to zero),
                       {\tt kt} is the parton transverse momentum $k_t$,
                       and {\tt mu} is the energy scale $\mu$ (in GeV). \\
  \midrule
  {\tt TMDpdf(kf,x,xbar,kt,mu, uval,dval,s,c,b,glu)}
                       & As the function above, but for hadron with flavour code {\tt kf}
                       ({\tt kf} = 2212 for proton and {\tt kf} =-2212 for antiproton)\\
  \midrule 
  {\tt TMDpdf(x,xbar,kt,mu)}
                       & Vector double-type function returning an array of $13$ variables 
                       with the values of $xF(x,{\bar x},\kt,\mu)$ ($F$ is the initialised uPDF/TMD):
                       at index $0,\dots,6$ is $\bar{t},\dots,\bar{d}$, at index $7$
                       is the gluon, and at index $8,\dots,13$ is $d,\dots,t$ densities for a proton target.\\
  \midrule
  {\tt TMDpdf(kf,x,xbar,kt,mu)}
                       & As the function above, but for hadron with flavour code {\tt kf}
                       ({\tt kf} = 2212 for proton and {\tt kf} =-2212 for antiproton)\\
  \midrule
  {\tt TMDpdf(x,xbar,kt,mu,xpq)}
                       & Void-type function filling an array of $13$ variables, {\tt xpq},
                       with the values of $xF(x,{\bar x},\kt,\mu)$ 
                       ($F$ is the initialised uPDF/TMD):
                       at index $0,\dots,6$ is $\bar{t},\dots,\bar{d}$, at index $7$
                       is the gluon, and at index $8,\dots,13$ is $d,\dots,t$ densities for a proton target.\\  
  \midrule
  {\tt TMDpdf(kf,x,xbar,kt,mu, xpq)}
                       & As the function above, but for hadron with flavour code {\tt kf}
                       ({\tt kf} = 2212 for proton and {\tt kf} =-2212 for antiproton)\\
   \bottomrule
  \end{tabularx}
 \caption{The function overload for {\tt TMDpdf} used to call uPDF/TMD parton sets.}
 \label{tab:TMDpdf}
\end{table}  
\begin{table}[!h]
\footnotesize
\centering
 \begin{tabularx}{\textwidth}{p{0.3\textwidth}X}
  \toprule
  Method & Usage \\
  \midrule
  {\tt TMDalphas(mu)} & Returns $\as$ used in the set initialised by {\tt TMDinit(name)}.\\
  \midrule
  {\tt TMDgetLam4( )} & Returns the value of $\Lambda_{QCD}$ at $N_f=4$.\\
  \midrule
  {\tt TMDgetNf( )}   & Returns the number of flavours, $N_f$, used for the computation of $\Lambda_{QCD}$.\\
  \midrule
  {\tt TMDgetOrderAlphaS( )}
                      & Returns the perturbative order of $\as$ used in the evolution of 
                      the TMD/uPDF set initialised by {\tt TMDinit(name)}.\\
  \midrule
  {\tt TMDgetOrderPDF( )} 
                      & Returns the perturbative order of the evolution of the TMD/uPDF 
                      set initialised by {\tt TMDinit(name)}.\\ 
  \midrule                  
  {\tt TMDgetXmin()}  & Returns the minimum value of the momentum fraction $x$ 
                      for which the TMD/uPDF set initialised by {\tt TMDinit(name)} was determined.\\
  \midrule
  {\tt TMDgetXmax()}  & Returns the maximum value of the momentum fraction $x$
                      for which the TMD/uPDF set initialised by {\tt TMDinit(name)} was determined.\\
  \midrule
  {\tt TMDgetQ2min()} & Returns the minimum value of the energy scale $\mu$ (in GeV) 
                      for which the TMD/uPDF set initialised by {\tt TMDinit(name)} was determined.\\  
  \midrule
  {\tt TMDgetQ2max()} & Returns the maximum value of the energy scale $\mu$ (in GeV) 
                      for which the TMD/uPDF set initialised by {\tt TMDinit(name)} was determined.\\
  \midrule
  {\tt TMDnumberPDF(name)} 
                      & Returns the identifier associated with 
                      the TMD/uPDF set initialised by {\tt TMDinit(name)}.\\                 
  \bottomrule
 \end{tabularx}
\caption{The list of methods included in the {\tt TMDutils.cc} file.}
\label{tab:tmdlibmethods}
\end{table}

The {\tmdlib} library is released together with the online
plotter platform {\tmdplotter},  available at
\url{http://tmdplotter.desy.de/}.
Two snapshots from a typical usage of 
{\tmdplotter} are shown in Fig.~\ref{fig:snapshot}: 
the gluon from the {\tt ccfm-JH-2013-set1} set is compared to the 
{\tt GBW} as a function of $k_t$ and $x$.

\clearpage

\section{Conclusions and feedback}
\label{sec:conclusions}

The authors of this manual set up a collaboration to develop and maintain {\tmdlib} and {\tmdplotter}, respectively a {\tt C++} library for handling different parameterisations of uPDFs/TMDs and a corresponding online plotting tool.
The redistribution of the fits has been agreed with the corresponding authors. 
The aim is to update these tools with more uPDF/TMD parton sets and new features, as they 
become available and are developed.

\section*{Acknowledgments}

The work of A.S. is part of the program of the Stichting voor Fundamenteel Onderzoek der Materie (FOM),
which is financially supported by the Nederlandse Organisatie voor Wetenschappelijk Onderzoek (NWO). 
E.R.N. acknowledges the kind hospitality of the theory group at the Nationaal instituut voor subatomaire fysica (Nikhef)
as part of the research activities within the ERC Advanced Grant program QWORK (contract no 320389).
T.R. is supported by the U.S. National Science Foundation under Grant No.\ PHY-0969739.

\bibliographystyle{heralhc} 
\raggedright 
\bibliography{ref}

\end{document}